\documentclass{article}
\usepackage[preprint]{spconfa4}
\usepackage{amsmath,graphicx}
\usepackage{cite}

\copyrightnotice{978-1-6654-6867-1/22/\$31.00~\copyright2022 IEEE}
                   
\title{Signal-informed DNN-based DOA estimation combining an external microphone and GCC-PHAT features}
\name{
    Ulrik Kowalk$^{1}$\thanks{This work was funded by the German Federal Ministry of Education and
Research under the funding program "Forschung an Fachhochschulen", Project ID: 13FH666IB6.}, 
    Simon Doclo$^{2}$, 
    and Joerg Bitzer$^{1}$
    }
\address{
    $^{1}$Jade University of Applied Sciences, Institute for Hearing Technology and Audiology, Oldenburg, Germany\\ 
    $^{2}$University of Oldenburg, Department of Medical Physics and Acoustics and\\Cluster of Excellence Hearing4all, Oldenburg, Germany
    }
\begin{document}
%
\maketitle
\begin{abstract}
Aiming at estimating the direction of arrival (DOA) of a desired speaker in a multi-talker environment using a microphone array, in this paper we propose a signal-informed method exploiting the availability of an external microphone attached to the desired speaker. The proposed method applies a binary mask to the GCC-PHAT input features of a convolutional neural network, where the binary mask is computed based on the power distribution of the external microphone signal. Experimental results for a reverberant scenario with up to four interfering speakers demonstrate that the signal-informed masking improves the localization accuracy, without requiring any knowledge about the interfering speakers.
\end{abstract}
\section{Introduction}
\label{sec:intro}
In the last decades, a wide range of direction of arrival (DOA) estimation methods using microphone arrays have been proposed, ranging from correlation-based approaches, e.g., exploiting the generalized cross-correlation (GCC) \cite{knapp1976generalized,1457990}, beamforming-based approaches, e.g., steered response power with phase transform (SRP-PHAT) \cite{dibiase2000high} or diagonal unloading \cite{salvati2020diagonal}, subspace-based approaches \cite{schmidt1986multiple} to approaches based on deep neural networks (DNNs) \cite{8170010,wang2018robust,8651493,zhang2019robust,mack2020signal,9054754}. A specific problem is the localization of a single desired speaker in a multi-talker scenario. Without any knowledge about the desired speaker, calculating a reliable DOA estimate is obviously a challenge \cite{sivasankaran2018keyword}. The sparse nature of speech -- temporally as well as spectrally -- may provide a valuable leverage point. Several authors have proposed the use of masking to guide the DOA estimation towards a desired speaker \cite{wang2018robust,zhang2019robust,mack2020signal}. 

\begin{figure}[t]
\vspace{-0.3cm}
    \includegraphics[width=\linewidth]{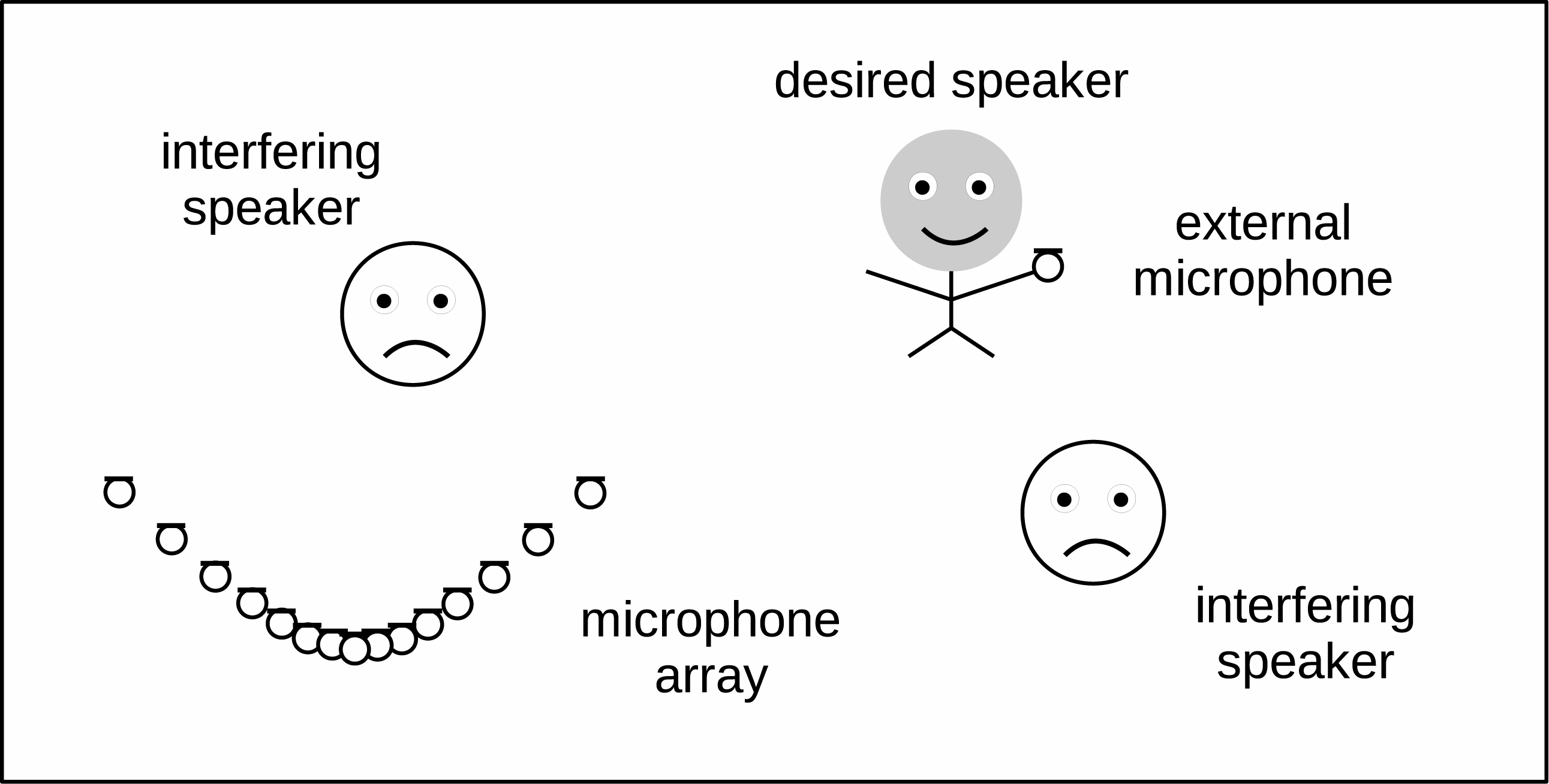}
    \caption{Scenario with one desired speaker and two interfering speakers. The external microphone is attached to the desired speaker.}
    \label{fig:Scenario}
\end{figure}

In this paper we assume the availability of an external microphone close to the desired speaker (see Fig.\,\ref{fig:Scenario}), which is exploited as an additional source of information (similarly as in \cite{farmani2017informed,fejgin2021comparison}). For example, one could consider a classroom scenario where the teacher (desired speaker) wears a microphone, e.g. to support a student with hearing disabilities, and the objective is to localize the teacher in the presence of multiple interfering speakers (students). 
We propose to utilize the power distribution from the external microphone signal to compute a binary mask that is applied to the input features of a DNN to calculate a signal-informed DOA estimate for the desired speaker. The algorithm is evaluated under reverberant conditions with a single desired speaker in a multi-talker environment. Results demonstrate that the signal-informed masking improves the localization accuracy, without requiring any knowledge about the acoustic scenario.

\vspace{-0.5cm}
\section{Signal Model}
We consider a reverberant scenario with one desired speaker, $J$ interfering speakers, and a small amount of background noise. The speech signals are captured by a microphone array with $M$ microphones and an external microphone close to the desired speaker (see Fig.\,\ref{fig:Scenario}). In the frequency domain, the desired speech signal is denoted by $D(\omega)$, while the $m$-th microphone signal and the external microphone signal are denoted by $Y_{m}(\omega)$ and $E(\omega)$, respectively. The $(M$+$1)$-dimensional vector of all microphone signals $\textbf{Y}(\omega) = [Y_{0}(\omega), Y_{1}(\omega),...,Y_{M-1}(\omega), E(\omega)]^{T}$, where $(\cdot)^{T}$ denotes transpose, can be written as 
\begin{equation}
    \textbf{Y}(\omega)=D(\omega)\textbf{H}_{D}(\omega) + \sum_{j=1}^{J}I_{j}(\omega)\textbf{H}_{j}(\omega) + \textbf{V}(\omega),
\end{equation}
where $\textbf{H}_{D}(\omega)$ denotes the vector of acoustic transfer functions between the desired speaker and the microphones, $I_{j}(\omega)$ denotes the $j$-th interfering speech signal, $\textbf{H}_{j}(\omega)$ denotes the vector of acoustic transfer functions between the $j$-th interfering speaker and the microphones and $\textbf{V}(\omega)$ denotes the background noise.
\section{Signal-informed DOA estimation exploiting an external microphone}
\begin{figure}[t]
    \includegraphics[width=\linewidth]{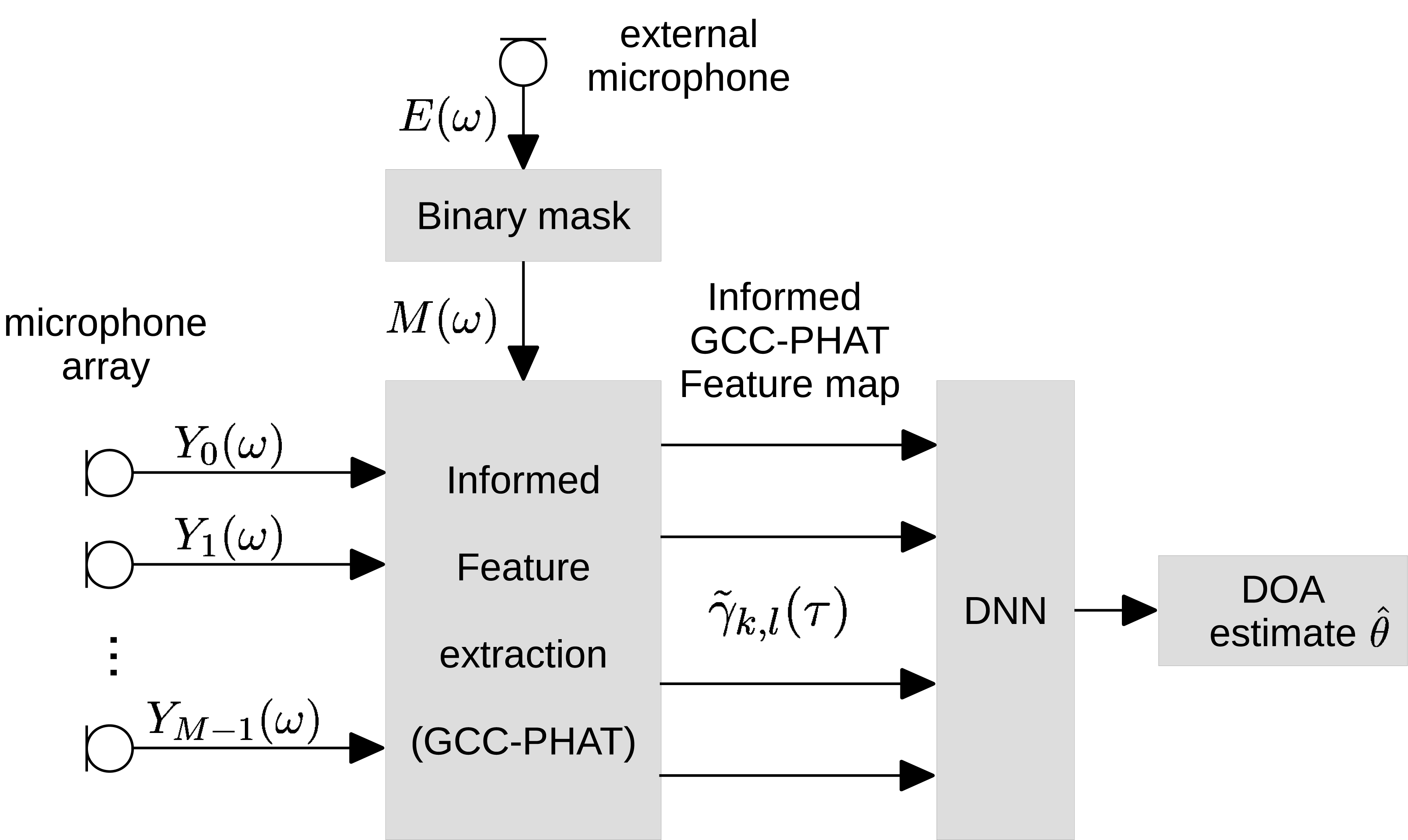}
    \caption{Block diagram of the signal-informed DOA estimation algorithm}
    \label{fig:Flow_Chart}
\end{figure}

The proposed algorithm uses time-domain generalized cross-correlation with phase transform (GCC-PHAT)\cite{knapp1976generalized} features as input to a convolutional neural network. We exploit the external microphone signal to generate a binary mask that is applied to the input features. The DOA estimation is formulated as a multi-class classification task with $C$=72 classes representing a set of DOAs with 5° resolution in the horizontal plane. For each audio frame a set of GCC-PHAT features is calculated and a DOA estimate is produced. Figure \ref{fig:Flow_Chart} depicts an overview of the algorithm. 
The remainder of this section is divided into three parts. In Section \ref{sec:InputFeature} the input features are described. In Section \ref{sec:InformedMasking} we explain how the external microphone signal can be used to compute a mask and guide the DOA estimation towards the desired speaker. Finally, in Section \ref{sec:DNN_Architecture}, the DNN architecture is presented.
\subsection{Input Features}
\label{sec:InputFeature}
As input features to the DNN we use the well-known GCC-PHAT. The GCC-PHAT between microphones $k$ and $l$ at time lag $\tau$ is defined as
\begin{equation}
    \gamma_{k,l}(\tau) = \mathcal{F}^{-1} \left\{ \frac{Y_{k}(\omega) \cdot Y^{*}_{l} (\omega)}{|Y_{k}(\omega) \cdot Y^{*}_{l} (\omega)|} \right\}\\
    \label{eq:GCC-PHAT_orig}
\end{equation}
where $(\cdot)^{*}$ denotes complex conjugation and $\mathcal{F}^{-1}$ denotes the inverse Fourier Transform. Through the PHAT weighting, the GCC-PHAT only depends on the phase difference between the microphone signals, i.e.
\begin{eqnarray}
    \gamma_{k,l}(\tau) &=& \mathcal{F}^{-1} \left\{ e^{i\phi_{k,l}(\omega)} \right\}\\
    &=& \mathcal{F}^{-1} \left\{ e^{i \cdot (\arg{\{Y_{k}(\omega)\}} - \arg{\{Y_{l}(\omega)\}})} \right\}
    \label{eq:GCC-PHAT_phase}
\end{eqnarray}
and not on the magnitude spectrum of the sound source. In practice, we limit the time lag $\tau$ to the interval [$-\tau_{max}$, $\tau_{max}-$1], where $\tau_{max}$ depends on the maximum inter-micro-phone distance of the array. Considering all possible microphone combinations (including $k$=$l$), a complete GCC-PHAT feature map has dimensions $M\times M\times(2\tau_{max})$. 
The fact that every source direction results in a distinct GCC-PHAT pattern inspires the usage of a DNN for image classification (see Section \ref{sec:DNN_Architecture}).
%
%
\subsection{Signal-informed Masking}
\label{sec:InformedMasking}
To guide the algorithm towards localizing the desired speaker in a multi-talker scenario, we propose to utilize the information from the external microphone. Since the external microphone is assumed to be close to the desired speaker, its power relative to the other speakers can be assumed to be large in the external microphone signal. We propose to generate a binary mask by comparing the magnitude of the external microphone signal $|E(\omega)|$ to a threshold, i.e.
\begin{equation}
    M(\omega) = \left\{
\begin{array}{ll}
1 & |E(\omega)| \geq P_{x}(|E(\omega)| ) \\
0 & \, \textrm{else} \\
\end{array}
\right. 
\label{eq:Binary_Mask}
\end{equation}
where $P_{x}(|E(\omega)|)$ denotes the $x$-th percentile of $|E(\omega)|$. It should be noted that to compute the mask $M(\omega)$, no knowledge about the interfering speakers nor the noise is required. Similar to \cite{mack2020signal}, where a binary mask was applied to raw input phases, we propose to apply the binary mask in (\ref{eq:Binary_Mask}) to the phase differences and add random noise, i.e.
\begin{multline}
    \tilde{\phi}_{k,l}(\omega) = M(\omega) \cdot (\arg{\{Y_{k}(\omega)\}} - \arg{\{Y_{l}(\omega)\}}) +\\(1 - M(\omega)) \cdot U(\omega),
    \label{eq:Mask_applied2}
\end{multline}
which are then used to compute an informed GCC-PHAT
\begin{equation}
    \tilde{\gamma}_{k,l}(\tau) = \mathcal{F}^{-1} \left\{e^{i \cdot \tilde{\phi}_{k,l}(\omega)} \right\}.
    \label{eq:Mask_applied1}
\end{equation}
Directly multiplying the mask with the phase differences, i.e. setting $x$ percent of the phase bins to zero, would result in a biased source direction. Therefore, uniformly distributed random noise $U(\omega)\in[0, 2\pi]$ is added in (\ref{eq:Mask_applied2}), which does not contribute to the DOA estimation but indirectly guides the estimation towards the spectral components dominated by $E(\omega)$. Because the mask is applied during feature generation, it may be changed without the need for retraining the DNN. Since the external microphone signal is transmitted directly, $E(\omega)$ is time-aligned with the microphone signals $Y_{0}(\omega)...Y_{M-1}(\omega)$ in order to account for the travel time of the sound wave from the source to the microphone array. We approximate this time difference as the time lag that maximizes the cross-correlation between $E(\omega)$ and the central microphone signal of the array.
\subsection{DNN architecture}
\label{sec:DNN_Architecture}
\begin{figure}[t]
    \includegraphics[width=\linewidth]{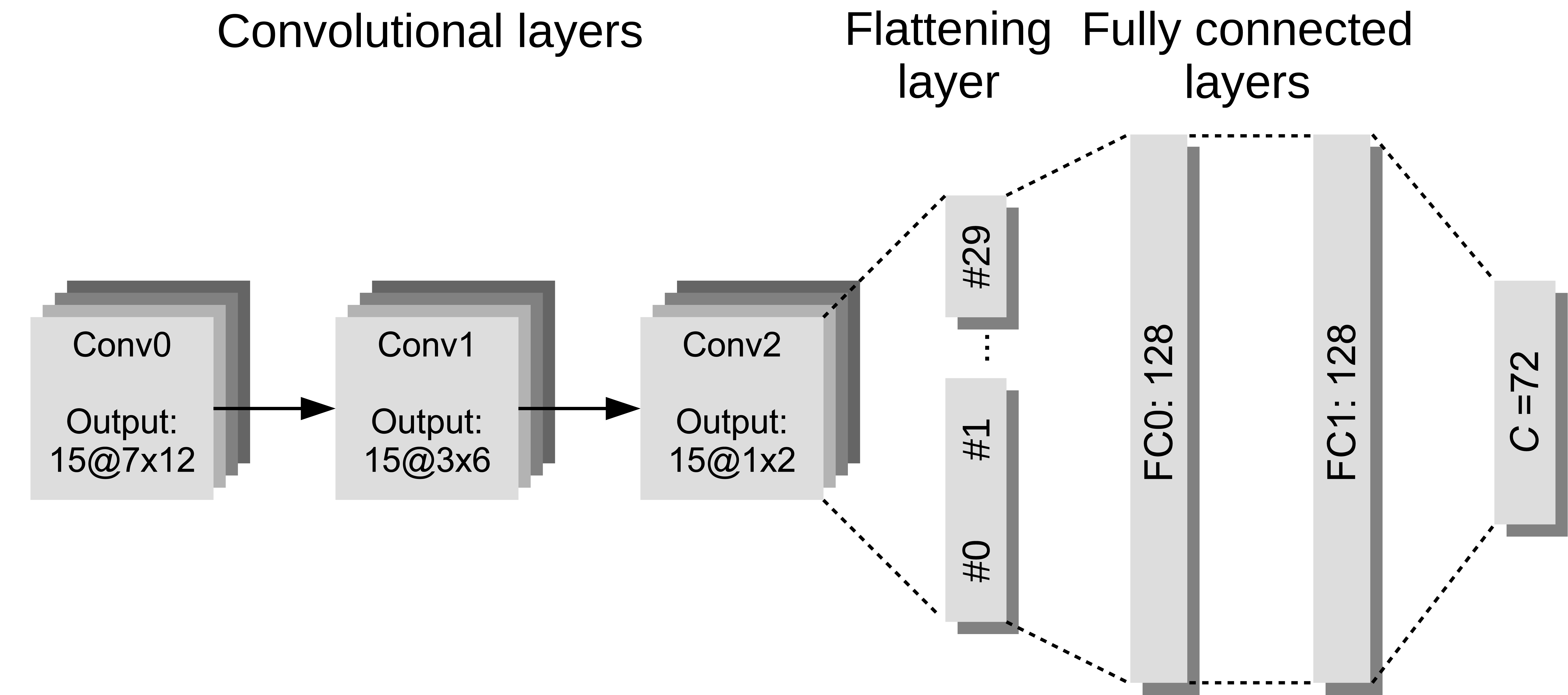}
    \caption{DNN architecture: 3 convolution layers with dimensionality reduction, 2 fully connected (FC) layers followed by an output layer with $C$=72 classes.}
    \label{fig:DNN_Architecture}
\end{figure}

Figure\,\ref{fig:DNN_Architecture} depicts the considered DNN architecture. Like in \cite{8170010} we use a cascade of 3 convolution layers, but with kernel dimensions of 3$\times$3. Each stage includes batch normalization, max pooling (2$\times$2 for the first two stages, 3$\times$3 for the last stage), 50\,\% dropout, and a leaky ReLU activation function. Through the continuous dimensionality reduction, the receptive field is gradually broadened, from small details in the feature (3$\times$3) to the whole feature in the end. After a flattening layer, two fully connected layers with 128 neurons are followed by the output layer, resulting in $C$=72 neurons, one for each class. A DOA estimate is defined as the class with the largest value at the output of the DNN. Comprising only 36008 learnable parameters, this architecture is comparatively small. 
\section{Experimental Validation}
\label{sec:ExperimentalValidation}
In this section we explain the acoustic setup and the training and evaluation of the proposed algorithm. It should be noted that training is only conducted using a single sound source and no masking is applied, whereas in the evaluation there is a desired speaker in a multi-talker environment and the binary mask is applied to guide the DOA estimation towards the desired speaker. 
\subsection{Acoustic Setup}
\label{sec:Setup}
For the experimental validation, we consider a non-uniformly spaced two-dimensional array with $M$=15 microphones, where the microphones are log-log-spaced on an arc, with a width of approximately 0.4\,m and a depth of approximately 0.13\,m (see Fig.\,\ref{fig:Scenario} for outline). According to the array geometry we assume $\tau_{max}$=12\,smpls. To simulate sound sources with directional cues, we convolve clean monophonic signals with room impulse responses (RIRs), that we generate using \textit{pyroomacoustics} \cite{scheibler2018pyroomacoustics}. All simulations are performed using Hann windowed non-overlapping frames with a length of 32\,ms at a sampling rate of 8\,kHz. 
\subsection{Training}
\label{sec:Training}
\raggedbottom
Training is conducted on single frames, each containing a single sound source. As already mentioned, no masking is applied during training, therefore the external microphone signal is not used. Half of the source signals for training consist of white noise, whereas the other half consist of speech. The speech signals are taken from the ``clean'' section of the \emph{LibriSpeech} corpus \cite{panayotov2015librispeech}, comprising 2700 recordings of male and female speakers. Aiming at achieving a good level of generalization against unseen conditions, we generate training data with a high intrinsic variance for all parameters. Every training signal contains a new set of RIRs with randomized room dimensions, position of the microphone array, position of the sound source in the room, as well as reverberation time. Finally, we add either white noise or babble noise with diffuse-like \cite{habets2006room} as well as spatially uncorrelated characteristics at different SNRs. All simulation parameters are given in Table\,\ref{tab:Variations}. The network is trained using the cross-entropy loss function, together with the Adam optimizer operating at a learning rate of $10^{-4}$ with mini-batches of 32 training samples and $10^{5}$ training samples per epoch.
\begin{table}[t]
    \centering
    \begin{tabular}{|l | l|}
        \hline
         Room dimensions: & [9.0, 5.0, 3.0]\,m $\pm$ [1.0, 1.0, 0.5]\,m \\ 
         Array position: & [4.5, 2.5, 1.5]\,m $\pm$ [0.5, 0.5, 0.5]\,m \\  
         Source distance: & 1.0 - 3.0\,m [within boundaries]\\
         Source direction: & 0°\ :\ 5°\ :\ 355°\\
         $T_{60}$: & 0.13\,s - 1.0\,s  \\
         SNR: & 0 - 30\,dB\\
         \hline
        \end{tabular}
    \caption{\label{tab:Variations}Simulation parameters}
\end{table}
\subsection{Evaluation}
\label{sec:Testing}
During evaluation, we only consider speech signals. Every scenario consists of one desired speaker and a set of interfering speakers $J\in[0,1,2,4]$, each interfering speaker having the same power as the desired speaker. For every $J$, the performance is evaluated for 5000 individual scenarios (trials), each 5\,s long. The evaluation data are generated with the same parameter variance as the training data (see Table\,\ref{tab:Variations}), except for the reverberation time fixed to 0.5\,s and the SNR fixed to 20\,dB. To prevent sources from overlapping, we impose a minimum angular distance of 5 classes ($\hat{=}25^{\circ}$) between the desired speaker and the closest interfering speaker, and at least one class ($\hat{=}5^{\circ}$) between two interfering speakers. The external microphone shares the same  coordinates as the desired speaker, but with an offset of 0.2\,m on the vertical axis. It is important to note, that in some scenarios, an interfering speaker may be much closer to the microphone array than the desired speaker, resulting in a negative signal-to-interference-ratio at the microphone array.
\subsection{Performance Measures}
\label{sec:Performance_Measures}
The DOA estimation performance is evaluated in terms of the absolute angular error (in degrees) as
\begin{equation}
    \delta = |\ \arg\left\{ e^{i2\pi\ \cdot\ (\hat{\theta}-\theta_{t})\ /\ 360^{\circ}}\right\}\ |\ \cdot\ \frac{360^{\circ}}{2\pi},
    \label{eq:Angular_Error}
\end{equation}
where $\hat{\theta}$ denotes the estimated DOA and $\theta_{t}$ denotes the ground truth DOA (both in degrees). 
A single DOA estimate is obtained per trial by computing the median of the DOA estimates over all frames that are labeled as speech. Speech frames are defined as frames whose energy in the external microphone signal is larger than 4\,dB below the global average. 
%
\section{Results}
\label{sec:Results}
In this section, we present the results for two sets of experiments. First, we investigate the influence of the proposed masking for a different number of interfering speakers $J$, using $P_{50}$ as the masking threshold. Second, we investigate the influence of the masking threshold $P_{x}$ for a different number of interfering speakers. The respective combination of $J$ and $P_{x}$ is referred to as ``condition''.

\begin{figure}[t]
    \includegraphics[width=\linewidth]{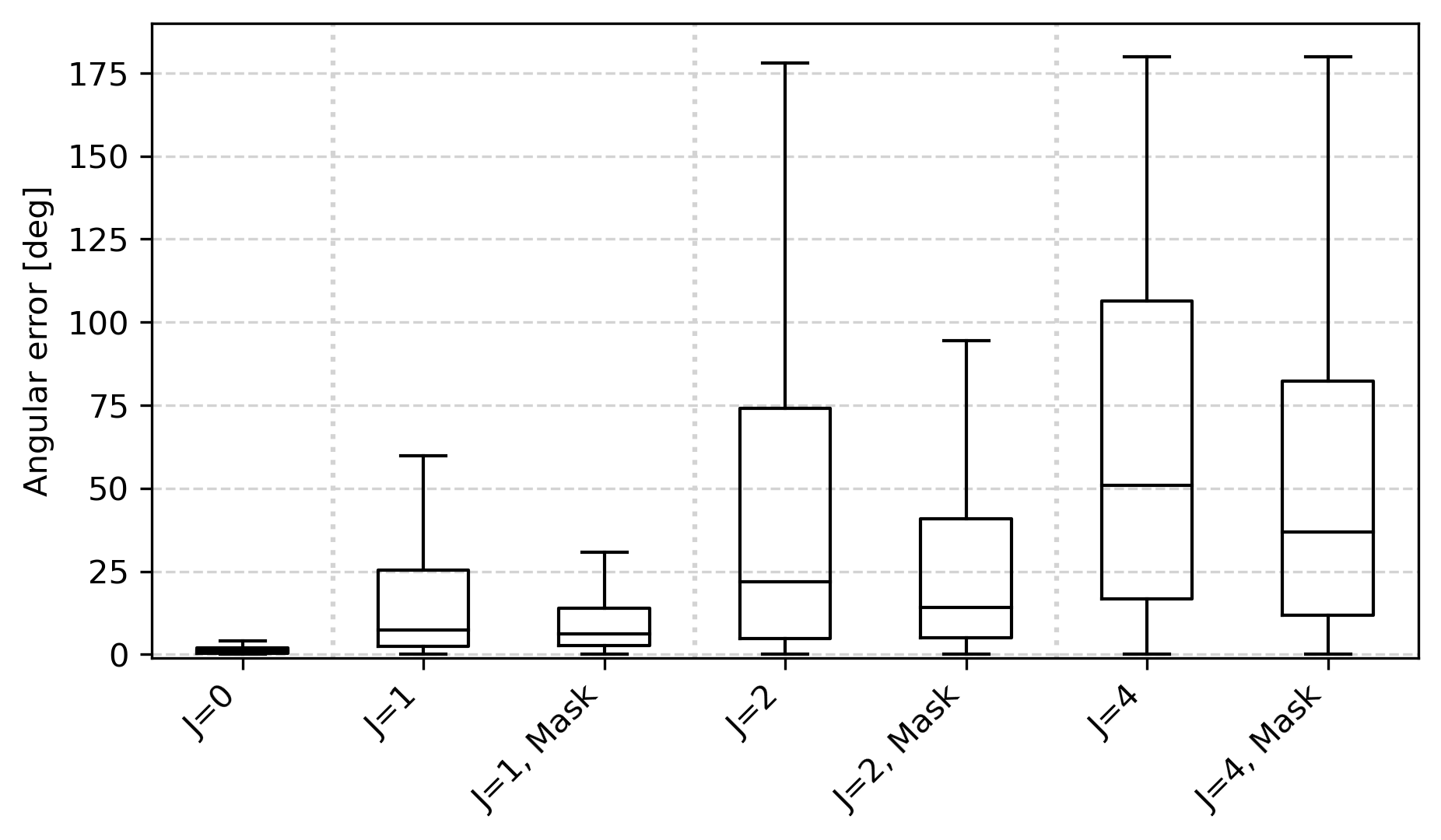}
    \caption{Angular error for different number of interfering speakers $J$, with and without masking}
    \label{fig:Results_cnn_gsc}
\end{figure}

Figure\,\ref{fig:Results_cnn_gsc} shows the impact of the proposed masking. For the condition with one interfering speaker, it can be observed that masking reduces the overall median error by about 14\,\%. For the conditions with 2 and 4 interfering speakers, masking reduces the overall median error by about 36\,\% and 28\,\%, respectively. In terms of the overall mean error, it can be observed that masking results in a reduction of about 45\,\% for the condition with one interfering speaker. For the conditions with 2 and 4 interfering speakers, masking reduces the overall mean error by about 30\,\% and 18\,\%, respectively.
Comprehensively, it can be observed that for all considered conditions, using the proposed source-informed features as input to the DNN substantially improves the localization accuracy of the desired speaker.

\begin{figure}[t]
    \includegraphics[width=\linewidth]{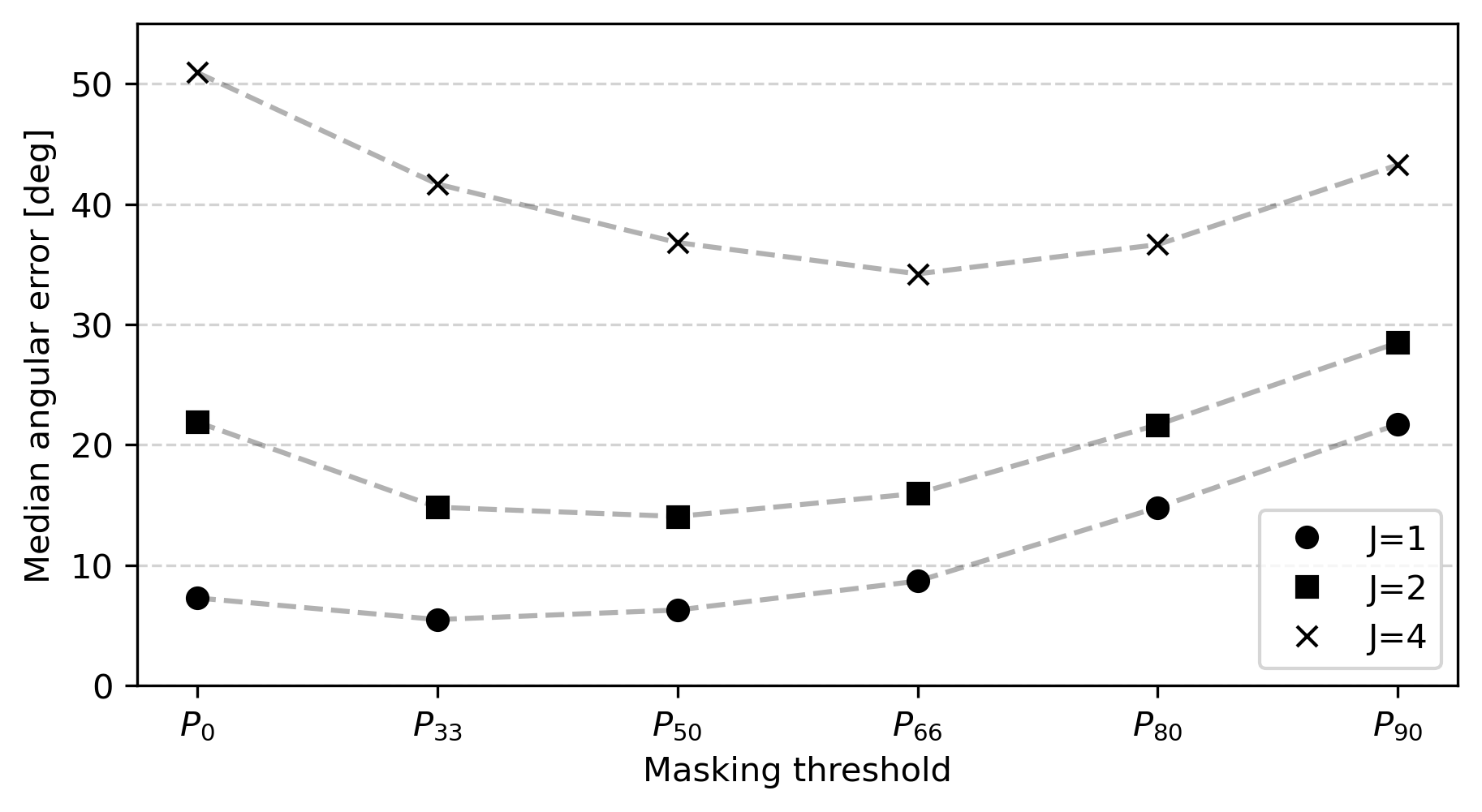}
    \caption{Impact of the masking threshold on the overall median angular error for different number of interfering speakers $J$}
    \label{fig:Impact_Of_Threshold}
\end{figure}

Figure\,\ref{fig:Impact_Of_Threshold} shows the impact of the masking threshold $P_{x}$. For scenarios with a higher number of interfering speakers, a larger threshold appears to yield a benefit. For the condition with one interfering speaker, $P_{33}$ presents itself as the optimal choice (overall median reduction of about 25\,\%). For the condition with 2 interfering speakers, where the desired source produces only about 33\,\% of the sound energy in the room, $P_{50}$ delivers the best results (overall median reduction of about 36\,\%), while for the condition with 4 interfering speakers, $P_{66}$ delivers the best results (overall median reduction of about 33\,\%). However, it can also be observed that a masking threshold between $P_{50}$ and $P_{66}$ yields good results for all considered acoustic scenarios.
\section{Conclusion}
\label{sec:Conclusion}
This paper has demonstrated the benefit from integrating an external microphone signal in a DNN-based DOA estimation algorithm where a desired speaker is to be localized in the presence of interfering speakers. The proposed algorithm applies a binary mask to the GCC-PHAT input features, where this mask is computed based on the power distribution of the external microphone signal. Experimental Results for a variety of acoustic scenarios show that the proposed algorithm significantly improves the DOA estimation accuracy without requiring any knowledge about the acoustic scenario. In future work we will investigate different ways of integrating the external microphone signal for informed DOA estimation.

%
%
%
\bibliographystyle{IEEEbib}
\bibliography{refs}

\begin{thebibliography}{10}

\bibitem{knapp1976generalized}
C.H. Knapp and G.C. Carter,
\newblock ``The generalized correlation method for estimation of time delay,''
\newblock {\em IEEE Trans. on Acoustics, Speech, and Signal Processing}, vol.
  24, no. 4, pp. 320--327, 1976.

\bibitem{1457990}
G.C. Carter,
\newblock ``Coherence and time delay estimation,''
\newblock {\em Proc. of the IEEE}, vol. 75, no. 2, pp. 236--255, 1987.

\bibitem{dibiase2000high}
J.H. DiBiase,
\newblock {\em A high-accuracy, low-latency technique for talker localization
  in reverberant environments using microphone arrays},
\newblock PhD thesis, Brown University, 2000.

\bibitem{salvati2020diagonal}
D.~Salvati, C.~Drioli, and G.L. Foresti,
\newblock ``Diagonal unloading beamforming in the spherical harmonic domain for
  acoustic source localization in reverberant environments,''
\newblock {\em IEEE/ACM Trans. on Audio, Speech, and Language Processing}, vol.
  28, pp. 2001--2012, 2020.

\bibitem{schmidt1986multiple}
R.~Schmidt,
\newblock ``Multiple emitter location and signal parameter estimation,''
\newblock {\em IEEE Trans. on Antennas and Propagation}, vol. 34, no. 3, pp.
  276--280, 1986.

\bibitem{8170010}
S.~Chakrabarty and E.A.P. Habets,
\newblock ``Broadband {DOA} estimation using convolutional neural networks
  trained with noise signals,''
\newblock in {\em Proc. IEEE Workshop on Applications of Signal Processing to
  Audio and Acoustics (WASPAA)}, 2017, pp. 136--140.

\bibitem{wang2018robust}
Z.-Q. Wang, X.~Zhang, and D.~Wang,
\newblock ``Robust speaker localization guided by deep learning-based
  time-frequency masking,''
\newblock {\em IEEE/ACM Trans. on Audio, Speech, and Language Processing}, vol.
  27, no. 1, pp. 178--188, 2018.

\bibitem{8651493}
S.~Chakrabarty and E.A.P. Habets,
\newblock ``Multi-speaker {DOA} estimation using deep convolutional networks
  trained with noise signals,''
\newblock {\em IEEE Journal of Selected Topics in Signal Processing}, vol. 13,
  no. 1, pp. 8--21, 2019.

\bibitem{zhang2019robust}
W.~Zhang, Y.~Zhou, and Y.~Qian,
\newblock ``Robust {DOA} estimation based on convolutional neural network and
  time-frequency masking,''
\newblock in {\em Proc. Interspeech}, 2019, pp. 2703--2707.

\bibitem{mack2020signal}
W.~Mack, U.~Bharadwaj, S.~Chakrabarty, and E.A.P. Habets,
\newblock ``Signal-aware broadband {DOA} estimation using attention
  mechanisms,''
\newblock in {\em Proc. IEEE International Conference on Acoustics, Speech and
  Signal Processing (ICASSP)}, 2020, pp. 4930--4934.

\bibitem{9054754}
R.~Varzandeh, K.~Adiloğlu, S.~Doclo, and V.~Hohmann,
\newblock ``Exploiting periodicity features for joint detection and {DOA}
  estimation of speech sources using convolutional neural networks,''
\newblock in {\em Proc. IEEE International Conference on Acoustics, Speech and
  Signal Processing (ICASSP)}, 2020, pp. 566--570.

\bibitem{sivasankaran2018keyword}
S.~Sivasankaran, E.~Vincent, and D.~Fohr,
\newblock ``Keyword-based speaker localization: Localizing a target speaker in
  a multi-speaker environment,''
\newblock in {\em Proc. Interspeech}, 2018, pp. 2703--2707.

\bibitem{farmani2017informed}
M.~Farmani, M.S. Pedersen, Z.-H. Tan, and J.~Jensen,
\newblock ``Informed sound source localization using relative transfer
  functions for hearing aid applications,''
\newblock {\em IEEE/ACM Trans. on Audio, Speech, and Language Processing}, vol.
  25, no. 3, pp. 611--623, 2017.

\bibitem{fejgin2021comparison}
D.~Fejgin and S.~Doclo,
\newblock ``Comparison of binaural {RTF}-vector-based direction of arrival
  estimation methods exploiting an external microphone,''
\newblock in {\em Proc. European Signal Processing Conference (EUSIPCO)}, 2021,
  pp. 241--245.

\bibitem{scheibler2018pyroomacoustics}
R.~Scheibler, E.~Bezzam, and I.~Dokmani{\'c},
\newblock ``Pyroomacoustics: A python package for audio room simulation and
  array processing algorithms,''
\newblock in {\em Proc. IEEE International Conference on Acoustics, Speech and
  Signal Processing (ICASSP)}, 2018, pp. 351--355.

\bibitem{panayotov2015librispeech}
V.~Panayotov, G.~Chen, D.~Povey, and S.~Khudanpur,
\newblock ``Librispeech: an {ASR} corpus based on public domain audio books,''
\newblock in {\em Proc. IEEE International Conference on Acoustics, Speech and
  Signal Processing (ICASSP)}, 2015, pp. 5206--5210.

\bibitem{habets2006room}
E.A.P Habets,
\newblock ``Room impulse response generator,''
\newblock {\em Technische Universiteit Eindhoven, Tech. Rep}, vol. 2, no. 2.4,
  pp. 1, 2006.

\end{thebibliography}
\end{document}